\documentclass[reprint,amsmath,amssymb,aps,showkeys,longbibliography,onecolumn]{revtex4-2}

\usepackage{graphicx}
\usepackage{dcolumn}
\usepackage{bm}

\usepackage{}	
\begin{document}

\title{Localization detection based on quantum dynamics}

\author{Kazue Kudo}
 \email{kudo@is.ocha.ac.jp}
\affiliation{%
Department of Computer Science, Ochanomizu University, Tokyo 112-8610, Japan
}%
\affiliation{%
Graduate School of Information Sciences, Tohoku University, Sendai 980-8579, Japan
}%

\date{\today}

\begin{abstract}
Detecting many-body localization (MBL) typically requires the calculation of high-energy eigenstates using numerical approaches. 
This study investigates methods that assume the use of a quantum device to detect disorder-induced localization.
Numerical simulations for small systems demonstrate how the magnetization and twist overlap, which can be easily obtained from the measurement of qubits in a quantum device, change from the thermal phase to the localized phase.
The twist overlap evaluated using the wave function at the end of the time evolution behaves similarly to the one evaluated with eigenstates in the middle of the energy spectrum under a specific condition.
The twist overlap evaluated using the wave function after time evolution for many disorder realizations is a promising probe for detecting MBL in quantum computing approaches.
\end{abstract}


\maketitle


\section{\label{sec:intro}Introduction}

Many-body localization (MBL) has recently attracted significant interest~\cite{alet2018,parameswaran2018,abanin2019,gopalakrishnan2020,tikhonov2021}.
MBL is a generalization of Anderson localization for disordered quantum many-body systems with interactions.
When the disorder is strong enough,
MBL prevents the system from thermalizing.
The transition from the thermal to the localized phase, i.e., MBL transition, is recognized as an eigenstate phase transition from the ergodic phase to the non-ergodic phase~\cite{pal2010,huse2014,pekker2017}.
The MBL transition is typically detected by quantities evaluated using eigenstates, such as entanglement, inverse participation ratio, and one-particle density matrix~\cite{bardarson2012,serbyn2013a,serbyn2013b,bera2015,bera2016,kjall2014,enss2017,orell2019,luitz2015,khemani2017a,khemani2017b,hopjan2020,zhang2018,gray2018}, and spectral properties, such as level statistics~\cite{pal2010,enss2017,luitz2015,khemani2017a,khemani2017b,kudo2018,orell2019}, as well as local observables~\cite{pal2010,enss2017,kjall2014,orell2019}.
The random matrix theory and quantum chaos conjecture support the relationship between spectral properties and quantum chaos~\cite{casati1980,bohigas1984} and also provide a framework for characterizing the MBL transition~\cite{dalessio2016,suntajs2020}.
Experimental studies have also captured the signature of the MBL transition in disordered quantum systems with different architectures~\cite{schreiber2015,kondov2015,smith2016,bordia2016,bordia2017,luschen2017a,luschen2017b,wei2018,xu2018,kohlert2019,rispoli2019,rubio2019,zhu2021,gong2021}.
In experiments, MBL is often explored by the investigation of quantum dynamics.
For example, the imbalance between the populations of even and odd sites in an atomic system is a measurable quantity for localization detection.
The initial state has populations only on even sites, while odd sites are empty.
Then, the imbalance is close to $1$ initially and relaxes to zero as the system thermalizes.
However, the imbalance maintains a finite value in the localized phase due to the initial state memory.

Recently, new approaches using quantum annealers were studied to simulate the properties of disordered quantum systems~\cite{king2018,harris2018,kairys2020,bando2020,bando2021,king2021a,king2021b,filho2022}.
Probing MBL is also within the scope of quantum annealers and quantum computers.
Although current quantum devices are still noisy and cannot compute exact eigenstates, some dynamical characteristics of MBL are robust against noise.
In the localized phase, local quantities are conserved to some extent.
For example, if the initial state is the all-spin-up state, the magnetization maintains a large value because of the initial state memory.
The magnetization, in this case, is similar to the imbalance in an atomic system in the sense that the memory effect characterizes localization.
A recent experiment using a quantum annealer detected the localization transition through magnetization measurements at the end of the time evolution~\cite{filho2022}.

This study investigated localization detection based on quantum dynamics in a disordered quantum spin chain.
Magnetization is a simple quantity that can be easily measured in quantum devices.
We also employed another measurable quantity, the twist overlap.
The twist overlap is a quantity proposed to detect the MBL transition and measures the extent to which an eigenstate overlaps with its twisted state~\cite{kutsuzawa2022}.
A twisted state is obtained by applying a twist operator that rotates spins over the chain at gradually increasing angles.
The twist overlap almost vanishes for thermal eigenstates, whereas it has a finite value for localized ones~\cite{kutsuzawa2022}.
The twist overlap can also be evaluated using the state after time evolution.
Using the twist overlap is convenient for localization detection in quantum devices because it can be easily obtained from the measurement of each spin.

The numerical simulations of small system sizes in this study demonstrate how the magnetization and twist overlap change from the thermal phase to the localized phase.
These quantities were obtained at the end of the time evolution and averaged over different disorder realizations.
They often oscillate at different frequencies for each disorder realization.
Thus, we also examined the time dependence of these quantities to understand their dynamics. 
The time dependence, whose observation in experiments may require enormous time and effort, is also helpful in understanding the characteristics of these quantities.
Although the difference between the thermal and localized phases in those quantities is evident, we cannot decide the existence of a phase transition from the limited numerical simulations.
This work aims to demonstrate the effectiveness of experimentally measurable quantities, specifically twist overlap, as a localization probe.

\section{\label{sec:method}Model and methods}

\subsection{Model}

The model used in this study is a one-dimensional transverse Ising model with local random fields, which is applied to a quantum annealer.
The Hamiltonian is given by
\begin{equation}
 H= \sum_{j=1}^{L-1}J_j\sigma^z_j\sigma^z_{j+1}+ \sum_{j=1}^L h_j\sigma^z_j
- \sum_{j=1}^L \Gamma_j\sigma^x_j,
\label{eq:H}
\end{equation}
where $L$ is the system size and $\sigma_j^x$ and $\sigma_j^z$ are the Pauli operators of components $x$ and $z$, respectively, at site $j$.
The local field $h_j$ consists of random numbers with a uniform distribution in the interval $[-w, w]$, where $w$ denotes the disorder strength. 
The interactions $J_j$ and transverse fields $\Gamma_j$ are given by $1+r_j$, where $r_j$ is a uniform random number in the interval $[-\sigma, \sigma]$.
Here, we refer to weak disorders in $J_j$ and $\Gamma_j$ as static noises.
They are introduced to mimic static noises of couplings between qubits and local fields in a quantum device.

\subsection{Entanglement entropy and twist overlap}

The MBL transition is typically detected using quantities calculated using eigenstates.
Moreover, the quantum dynamics reflects the properties of eigenstates.
Before examining the quantum dynamics, we examine several quantities calculated using eigenstates.
Here, we employ the half-chain entanglement entropy and twist overlap evaluated with eigenstates in the middle of the energy spectrum.  

Half-chain entanglement entropy is defined by
\begin{equation}
 S_E=-\mathrm{Tr} \rho_A\log\rho_A,
\end{equation}
where $\rho_A$ is the reduced density matrix for subsystem $A$.
The subsystem $A$ corresponds to the first half of the spin chain.
Eigenstates obey the volume and area laws of entanglement in the thermal and localized phases, respectively.
In other words, $S_E$ decreases when the thermal-to-MBL transition occurs.
The transition point is characterized by the variance peak of the half-chain entanglement entropy~\cite{kjall2014,khemani2017a}.

The twist operator is defined by
\begin{equation}
 U_{\rm twist}=\exp\left[ \frac{i}{2}\sum_{j=1}^L\theta_j\sigma^z_j\right].
\end{equation}
It generates a spin-wave-like excitation by rotating the spins around the $z$ axis at angles $\theta_j=2\pi j/L$~\cite{nakamura2002,kutsuzawa2022}.
The factor $1/2$ originates from spin $\frac12$.
The overlap between a state $|\psi\rangle$ and its twisted state $U_{\rm twist}|\psi\rangle$ is the twist overlap, which is represented as
\begin{equation}
 z=\langle\psi|U_{\rm twist}|\psi\rangle,
\end{equation}
where $|\psi\rangle$ denotes the eigenstate of the Hamiltonian in the original definition~\cite{kutsuzawa2022}.
In the thermal phase, the twist overlap is expected to vanish because the twisted state with a spin-wave-like excitation is orthogonal to the original state.
In contrast, the long-wavelength perturbation given by the twist operator has little effect on the eigenstates in the localized phase, which implies a finite twist overlap.

In quantum devices, the twist overlap is easily obtained by measuring each qubit.
Writing $|\psi\rangle$ as a linear combination $|\psi\rangle=\sum_s\alpha_s|s\rangle$ of the computational basis $\{|s\rangle\}$ yields
\begin{equation}
 z=\sum_{s,s'}\alpha_s\alpha_{s'}^*\exp
\left[ \frac{i}{2}\sum_{j=1}^L\theta_j s^z_j\right]\langle s'|s\rangle
=\sum_s|\alpha_s|^2\exp\left[ \frac{i}{2}\sum_{j=1}^L\theta_j s^z_j\right],
\end{equation}
where $\sigma^z_j|s\rangle=s^z_j|s\rangle$.
Since each measurement provides the configuration of $s_j^z$ ($j=1,\ldots, L$) with probability $|\alpha_s|^2$, many measurements provide the expected value of the twist overlap.

\subsection{Time evolution}

The solution of the Schr{\"o}dinger equation,
\begin{equation}
 i\frac{d}{dt}|\psi(t)\rangle
=H|\psi\rangle,
\end{equation}
is represented as
\begin{equation}
 |\psi(t)\rangle = \exp(-iHt)|\psi_0\rangle,
\label{eq:psi1}
\end{equation}
where $|\psi_0\rangle$ denotes the initial state.
In the numerical simulations below, the initial state is the all-spin-up state.
The exact diagonalization of the Hamiltonian provides the time dependence of $|\psi(t)\rangle$.
We observe the magnetization and twist overlap at the end of the time evolution,which is at the final time $t=T_{\rm fin}$.

When the initial state is expressed as $|\psi_0\rangle=\sum_k c_k|\phi_k\rangle$ with eigenstates $|\phi_k\rangle$ of the Hamiltonian, Eq.~\eqref{eq:psi1} becomes
\begin{equation}
 |\psi(t)\rangle = \sum_{k=1}^{2^L} c_k\exp(-iE_kt)|\phi_k\rangle,
\label{eq:psi2}
\end{equation}
where $E_k$ denotes the eigenenergy corresponding to $|\phi_k\rangle$.
The $z$ component of the magnetization defined by $M_z\equiv\langle\psi|\sum_j\sigma^z_j|\psi\rangle$ evolves as
\begin{equation}
 M_z(t)=\sum_{k,l=1}^{2^L}c_kc_l^* 
\langle\phi_l|\sum_{j=1}^L\sigma^z_j|\phi_k\rangle
e^{-i(E_k-E_l)t}.
\label{eq:Mz_t}
\end{equation}
Similarly, the twist overlap evolves as
\begin{equation}
 z(t)=\sum_{k,l=1}^{2^L}c_kc_l^* 
\langle\phi_l|U_{\rm twist}|\phi_k\rangle
e^{-i(E_k-E_l)t}.
\label{eq:z_t}
\end{equation}
Equations~\eqref{eq:Mz_t} and \eqref{eq:z_t} suggest the oscillatory behavior of $M_z(t)$ and $z(t)$.
Therefore, the observed quantities depend on both the eigenstates and the final time.
Since the frequency and amplitude of the oscillation differ from sample to sample, the average over disorder realizations characterizes the observed quantities if the final time is sufficiently large.

\section{\label{sec:result}Results}

\subsection{\label{sec:eigen}Quantities calculated from eigenstates}

Before investigating quantum dynamics, we confirm localization properties characterized by eigenstates of the Hamiltonian.
They help understand quantities obtained from quantum dynamics.
The exact diagonalization of the Hamiltonian without noise ($\sigma=0$) was performed for each pair of the system size $L$ and disorder strength $w$. 
The number of disorder realizations was $10^4$ for $L=8$ and $10^3$ for $L=10$ and $12$.
For each realization, the half-chain entanglement entropy and twist overlap were calculated and averaged over 20 eigenstates around the center of the energy spectrum.

\begin{figure}[tbh]
\centering
\includegraphics[width=0.4\textwidth]{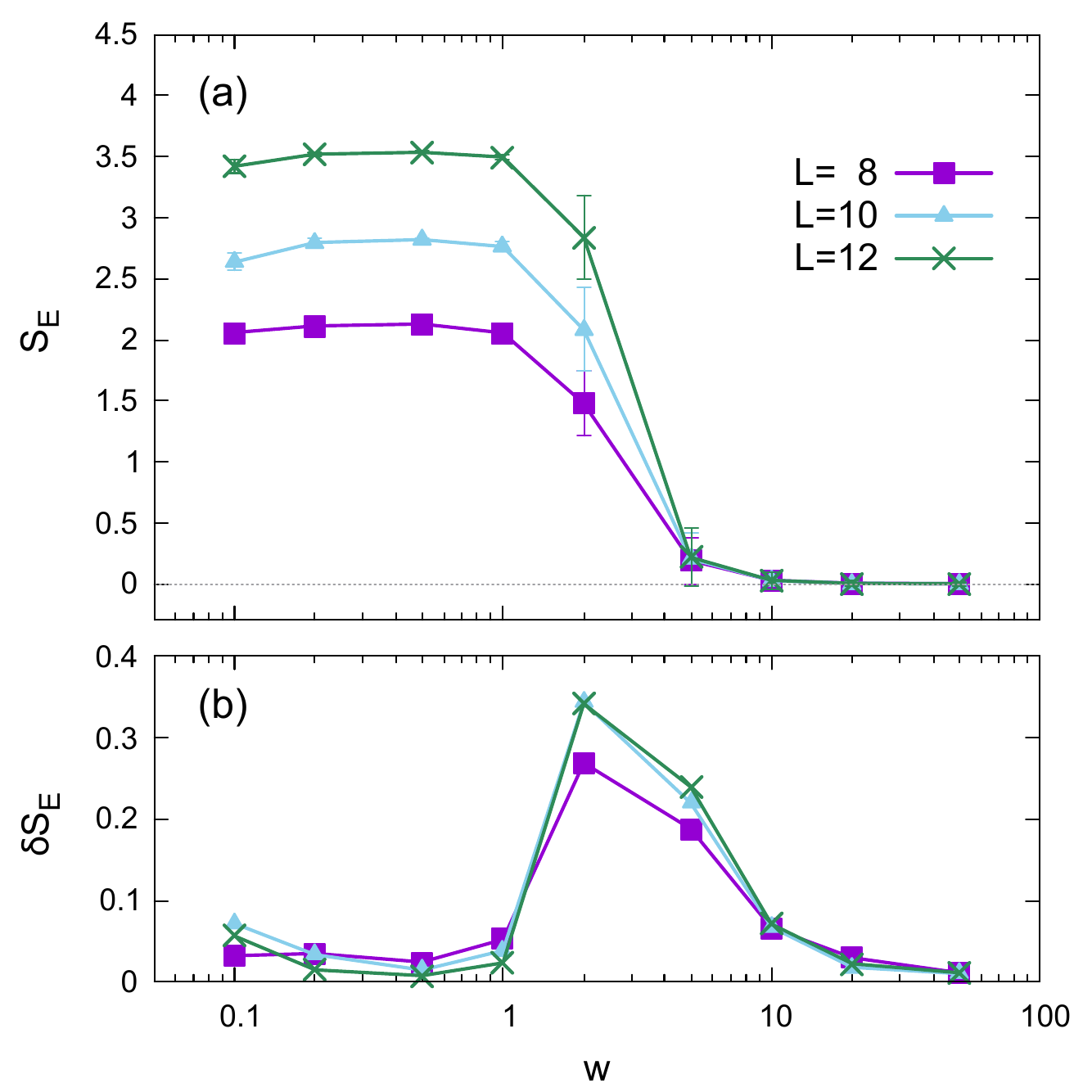}
\includegraphics[width=0.4\textwidth]{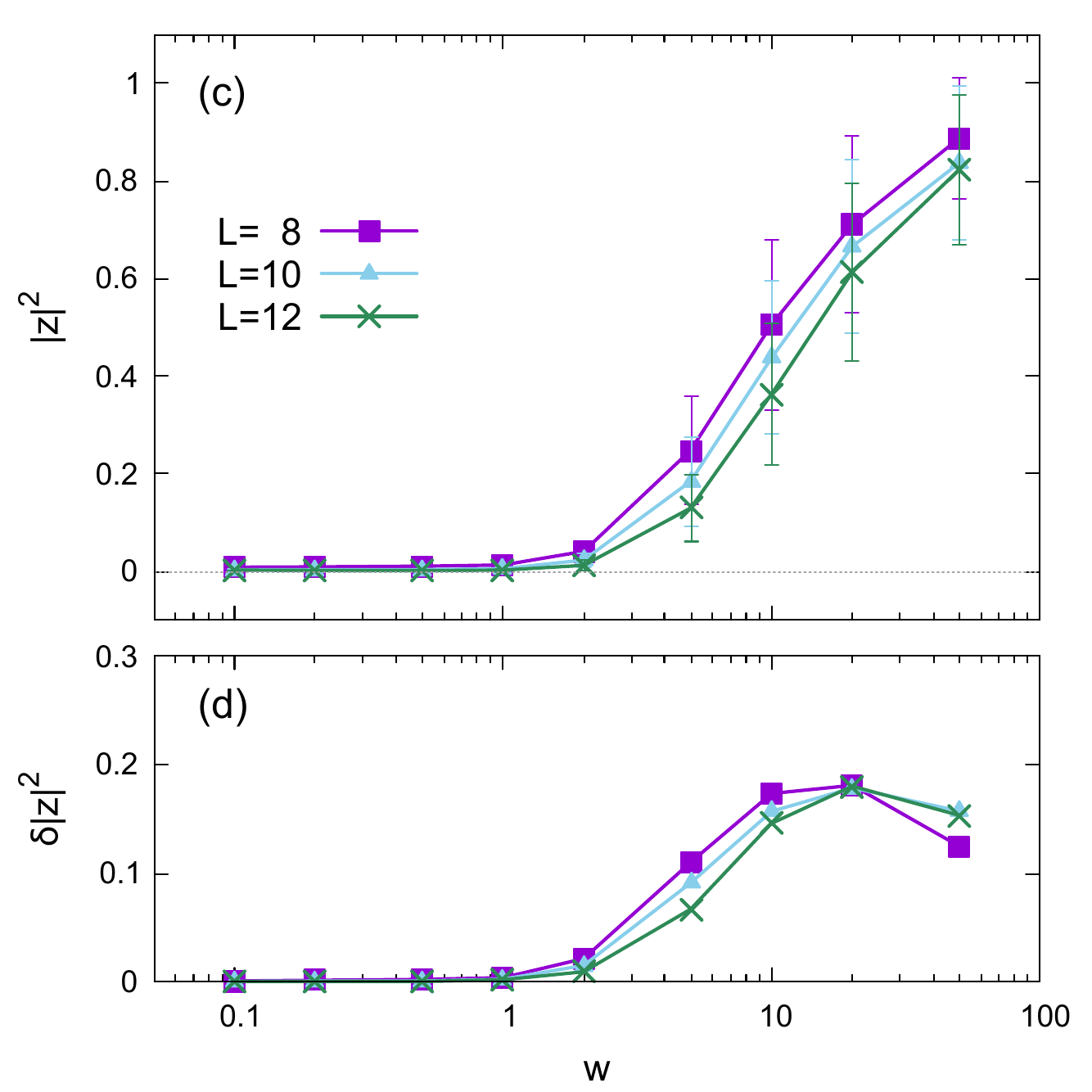}
\caption{\label{fig:eig}%
Disorder-strength dependence of the entanglement entropy (a) and (b) and the twist overlap (c) and (d), calculated using eigenstates for different system sizes in the noiseless case ($\sigma=0$).
(a) and (b) plot the half-chain entanglement entropy $S_E$ and its standard deviation $\delta S_E$ as functions of disorder strength $w$, respectively.
Similarly, (c) and (d) plot the absolute square of the twist overlap $|z|^2$ and its standard deviation $\delta|z|^2$, respectively.
The error bars in (a) and (c) represent standard deviation. 
}
\end{figure}

Figure~\ref{fig:eig} shows the dependence on the disorder strength $w$ of the half-chain entanglement entropy $S_E$ and the absolute square of the twist overlap $|z|^2$ averaged over the disorder realizations.
The averages of $S_E$ and $|z|^2$ are plotted with error bars in Figs.~\ref{fig:eig}(a) and (c), respectively, where the error bar represents the standard deviation.  
Figures~\ref{fig:eig}(b) and (d) show the standard deviations of the half-chain entanglement entropy and the absolute square of the twist overlap, respectively.

The disorder-strength dependence of the half-chain entanglement entropy shown in Figs.~\ref{fig:eig}(a) and (b) shows behavior similar to that of Ref.~\cite{filho2022}, although the values were different because of the differences in the models.
For each $L$, the variance (standard deviation) peaks around $w\simeq 1$--$5$, where the transition or crossover between the thermal and localized phases occurs.

As expected, the twist overlap, whose absolute square is shown in Fig.~\ref{fig:eig}(c), increases with the disorder strength.
The variance is almost zero in the weak-disorder region and becomes finite in the strong-disorder region.  
These results indicate that both the average and variance of the twist overlap are almost zero in the thermal phase, but large in the localized phase.
The disorder-strength dependence of the variance was not observed in Ref.~\cite{kutsuzawa2022}, which used the random-field Heisenberg chain.
The difference in the variance behavior is likely due to differences in the models.

Whereas the peaks of $\delta S_E$ appear in the middle of changes in $S_E$, those of $\delta|z|^2$ occur as $|z|^2$ is large enough.
The peaks of $\delta|z|^2$ appear probably due to the saturation of $|z|^2$. 
Thus, the peak position of $\delta|z|^2$ is not related to the transition or crossover between the thermal and localized phases.

\subsection{\label{sec:qd}Properties based on quantum dynamics}

In this subsection, we investigate the magnetization and twist overlap evaluated using the wave function at the end of time evolution.
The wave function $|\psi(t)\rangle$ at the final time $t=T_{\rm fin}$ was calculated from the exact diagonalization of the Hamiltonian.
The final time was $T_{\rm fin}=10$, which is sufficient to capture the difference between the thermal and localized phases, as shown in the following subsection.
The initial state is taken as the all-spin-up state.
$M_z=\langle\psi(T_{\rm fin})|\sum_j\sigma^z_j|\psi(T_{\rm fin})\rangle$
and
$z=\langle\psi(T_{\rm fin})|U_{\rm twist}|\psi(T_{\rm fin})\rangle$
were calculated for each disorder realization.
The number of disorder realizations was the same as that in the previous subsection: 
$10^4$ for $L=8$ and $10^3$ for $L=10$ and $12$.

\begin{figure}[tbh]
\centering
\includegraphics[width=0.4\textwidth]{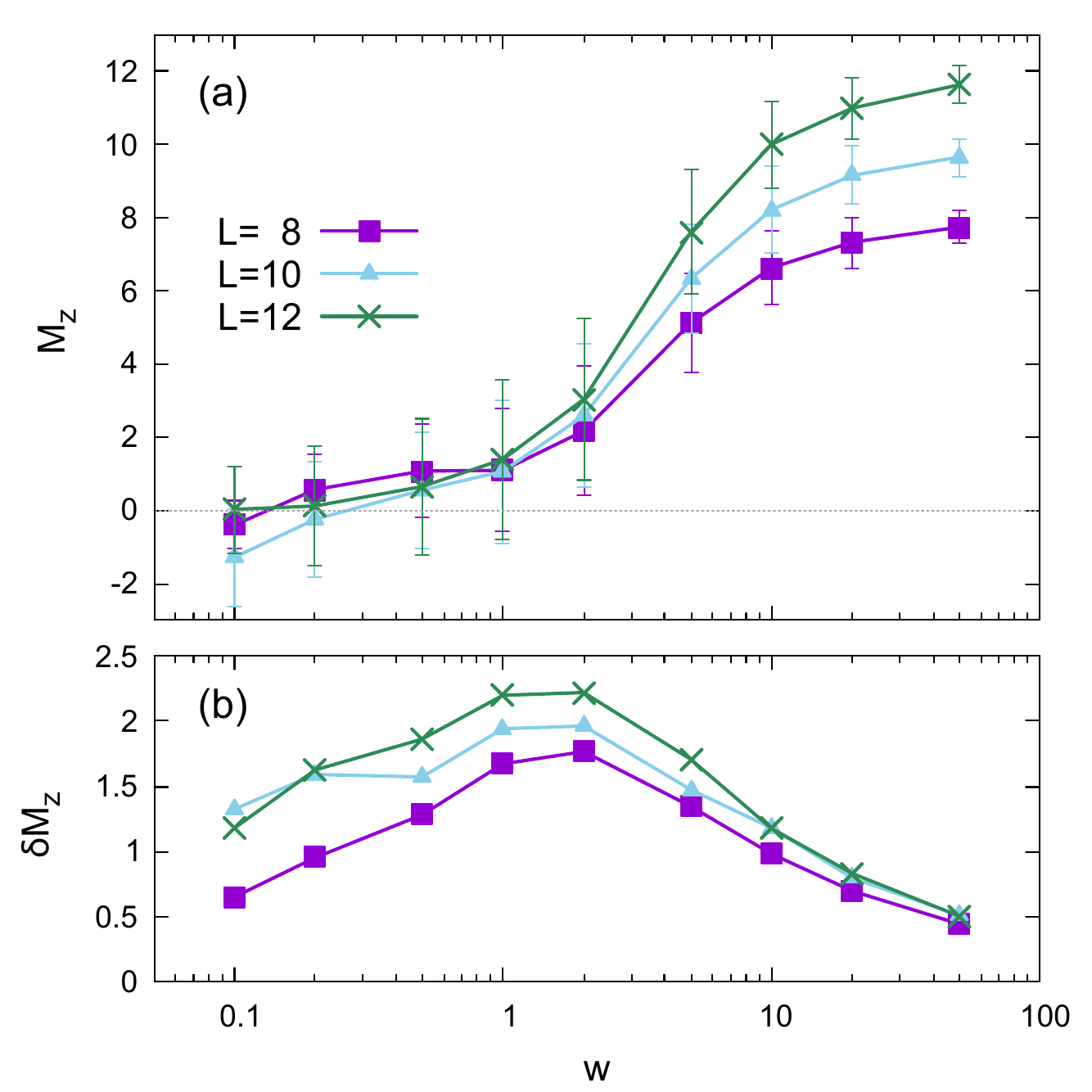}
\includegraphics[width=0.4\textwidth]{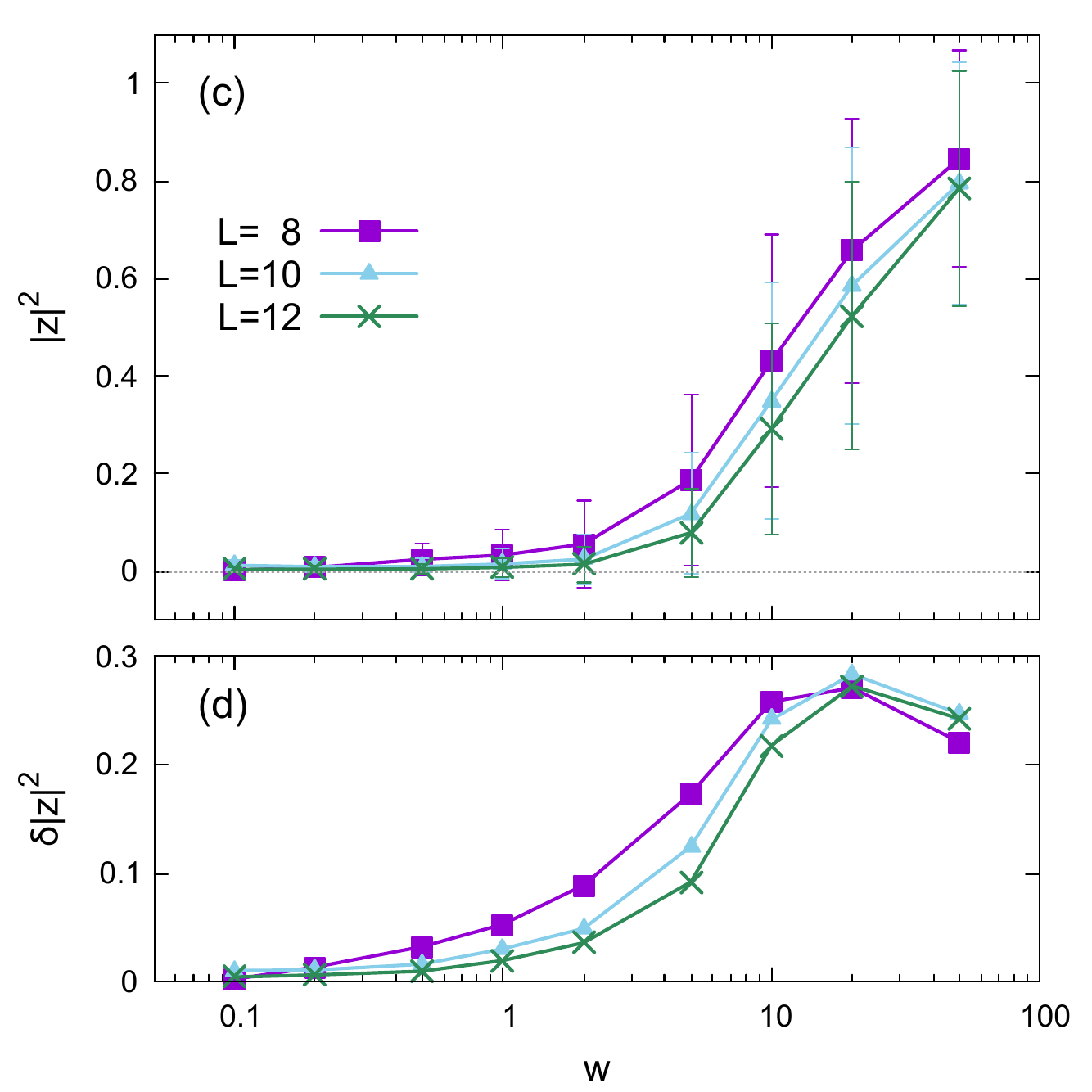}
\caption{\label{fig:dyn}%
Disorder-strength dependence based on quantum dynamics for different system sizes in the noiseless case ($\sigma=0$).
(a) and (b) plot the $z$ component of the magnetization $M_z$ and its standard deviation $\delta M_z$ as functions of disorder strength $w$, respectively.
(c) and (d) plot the absolute square of the twist overlap $|z|^2$ and its standard deviation $\delta|z|^2$, respectively.
The error bars in (a) and (c) represent standard deviation. 
}
\end{figure}

Figure~\ref{fig:dyn} shows the dependence on the disorder strength $w$ of the $z$ component of the magnetization $M_z$ and the absolute square of the twist overlap $|z|^2$ averaged over the disorder realizations.
Here, the noise strength is $\sigma=0$.
The averages of $M_z$ and $|z|^2$ are plotted with error bars in Figs.~\ref{fig:dyn}(a) and (c), respectively, where the error bar represents standard deviation.  
Figures~\ref{fig:dyn}(b) and (d) show the standard deviations of the $z$ component of the magnetization and the absolute square of the twist overlap, respectively.
 
The average magnetization is $M_z\simeq 0$ in the weak-disorder region, indicating thermalization.
When the disorder is strong enough, $M_z\simeq L$, which is a signature of the memory effect because $M_z=L$ in the initial state.
The memory effect is characteristic of the localized phase, which was also observed in Ref.~\cite{filho2022}.
The variance (standard deviation) of the magnetization peaks at a slightly weaker disorder strength than that of the entanglement entropy.
Since the magnetization fluctuates with time and can have negative values, $\delta M_z$ is relatively large in the weak-disorder region. 
Thus, the variance peak of the magnetization in this situation cannot apply to determining the transition or overlap point.

The twist overlap also increases with the disorder strength.
However, the memory effect is not the leading cause for the large value of $|z|^2$ in the strong-disorder region.
If the memory effect dominates the twist overlap behavior, $|z|^2$ should be close to $1$, and its variance should be small.
Considering that the twist overlap shown in Fig.~\ref{fig:dyn}(c) is similar to that in Fig.~\ref{fig:eig}(c), we expect that the behavior of the twist overlap reflects the properties of eigenstates.
The variance (standard deviation) in Fig.~\ref{fig:dyn}(d) is relatively large compared with that in Fig.~\ref{fig:eig}(d).
The large variance in Fig.~\ref{fig:dyn}(d) is due to the difference in the eigenstates of different disorder realizations and the oscillatory behavior of $z(t)$.

\begin{figure}[tbh]
\centering
\includegraphics[width=0.4\textwidth]{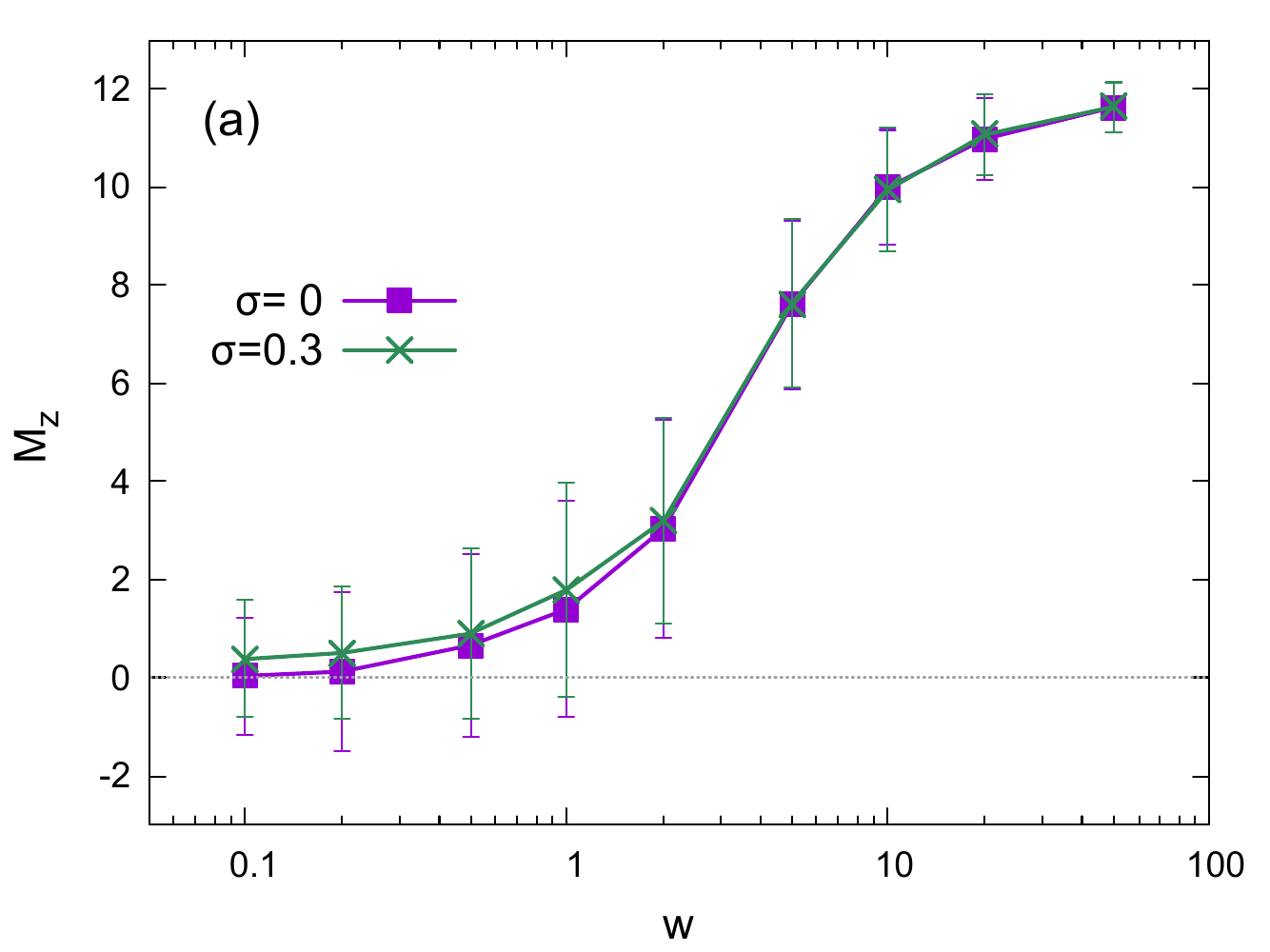}
\includegraphics[width=0.4\textwidth]{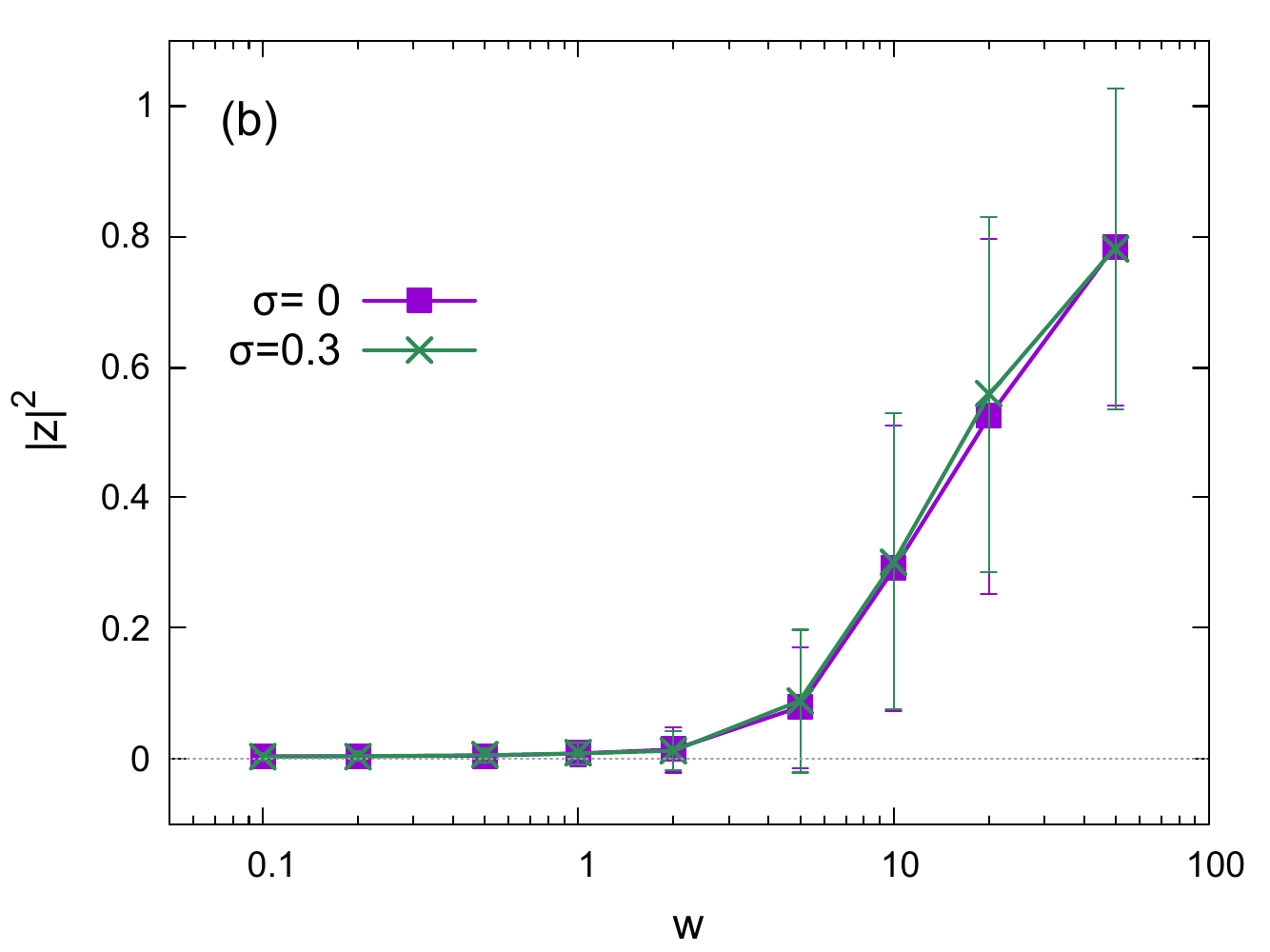}
\caption{\label{fig:dis}%
Disorder-strength dependence of (a) the $z$ component of the magnetization and (b) the absolute square of the twist overlap $|z|^2$ for different noise strengths.
The error bars represent the standard deviation.
The system size is $L=12$.
}
\end{figure}

Similar results to the noiseless case ($\sigma=0$) also appear in the presence of static noise.
While several types of noises exist in quantum devices, we here consider static noises in the interaction between spins and the transverse field.
As shown in Fig.~\ref{fig:dis}, the disorder strength of $M_z$ and $|z|^2$ has little dependence on noise strength $\sigma$.
However, the time evolution is affected by the static noises, as shown in the following subsection.
The details of the time evolution are averaged out in the results in Fig.~\ref{fig:dis}, making the noise dependence negligible.

\subsection{\label{sec:td}Time dependence}

\begin{figure}[tbh]
\centering
\includegraphics[width=\textwidth]{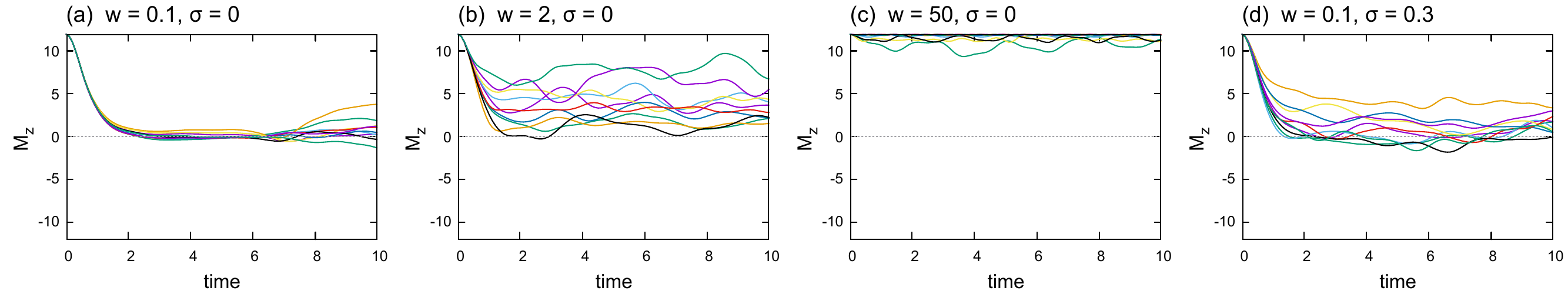}
\includegraphics[width=\textwidth]{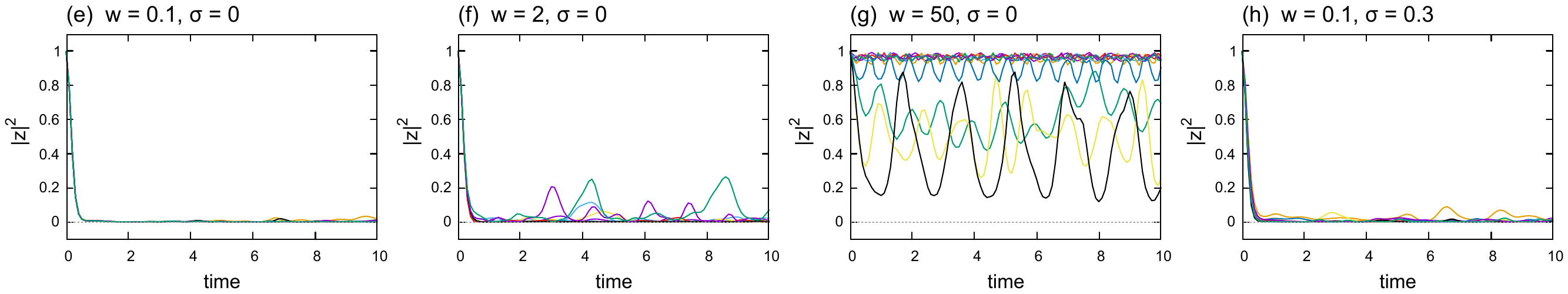}
\caption{\label{fig:time}%
Time dependence of (a)--(d) the magnetization and (e)--(h) the absolute square of the twist overlap.
Each graph plots ten different samples of the time series.
(a) and (e) share the same wave-function samples, and the same applies to (b) and (f), (c) and (g), and (d) and (h). The system size is $L=12$.
}
\end{figure}

The time dependence of the magnetization and twist overlap helps us to understand the characteristics of quantum dynamics in the system.
Figure~\ref{fig:time} illustrates $M_z(t)$ (in the upper row) and $|z(t)|^2$ (in the lower row) for several combinations of disorder strength $w$ and noise strength $\sigma$.
Each graph plots ten samples of the time series calculated at different disorder realizations.
The graphs in the same column, for example, (a) and (e), share the same wave-function samples.
That is, the curves of $M_z(t)$ and $|z(t)|^2$ with the same color in the same column are evaluated using the same wave functions.

Figures~\ref{fig:time}(a)--(c) and (e)--(g) demonstrate how the behaviors of $M_z(t)$ and $|z(t)|^2$ change with disorder strength $w$ in the noiseless case ($\sigma=0$).
Initially, $M_z=L=12$, which then decreases to $M_z\simeq 0$ when the disorder is weak.
As the disorder strengthens, $M_z(t)$ fluctuates around positive values.
Finally, $M_z(t)$ fluctuates around the initial value in the localized phase. 
However, $|z|^2=1$ at the initial time decreases rapidly to $|z|^2\simeq 0$ when the disorder is weak.
As the disorder strengthens, the fluctuations in $|z|^2$ become noticeable.
In contrast to $M_z(t)$, some samples of $|z(t)|^2$ oscillate with large amplitudes around relatively low values.

Figures~\ref{fig:time}(d) and (h) show the time dependence of $M_z(t)$ and $|z(t)|^2$, respectively, for the noisy ($\sigma=0.3$) and weak-disorder ($w=0.1$) cases. 
A comparison of Figs.~\ref{fig:time}(a) and (d) suggests that the time average of $M_z(t)$ is larger in the noisy case than that in the noiseless case in general.
The difference in time dependence indicates that the static noise in the interaction strength and transverse field affects the quantum dynamics, which is not reflected in Fig.~\ref{fig:dis}.

\section{\label{sec:discussion}Discussion}

The absolute square of the twist overlap $|z|^2$ exhibits a slight variance in the weak-disorder region, as shown in Figs.~\ref{fig:dyn} and \ref{fig:dis}.
However, the $z$ component of the magnetization $M_z$ has a relatively large variance in the same region.
The time dependences of $M_z(t)$ and $|z(t)|^2$ also support this behavior, which seems curious.
Figures~\ref{fig:eig}(c) and (d) illustrate that $|z|^2\simeq 0$ in the small-disorder region, implying that $\langle\phi_l|U_{\rm twist}|\phi\rangle$ in Eq.~\eqref{eq:z_t} nearly vanishes in the middle of the energy spectrum.
Since the eigenstates in the high and low regions of the energy spectrum also contribute to the time-dependent $|z(t)|^2$, the variance of $|z|^2$ has a small finite value in Figs.~\ref{fig:dyn} and \ref{fig:dis}.
However, $M_z(t)$ fluctuates around zero because of thermalization, which causes a relatively large variance.

As shown in Fig.~\ref{fig:time}, $|z(t)|^2$ oscillates with a large amplitude in some strong-disorder cases, even though $M_z(t)$ remains around the initial value.
Large-amplitude oscillations arise from the combination of $\langle\phi_l|U_{\rm twist}|\phi_k\rangle$ with different eigenstates $|\phi_l\rangle$ and $|\phi_k\rangle$. 
As shown in Figs.~\ref{fig:eig}(c) and (d), the average and variance of $|z|^2$ are significant in the strong-disorder region, which supports the variation in the combination of $\langle\phi_l|U_{\rm twist}|\phi_k\rangle$.

\section{\label{sec:conc}Conclusions}

We investigated a method that assumes the use of a quantum device to detect disorder-induced localization.
Localization in a disordered spin chain is detected by evaluating the magnetization and twist overlap at the end of the time evolution for many disorder realizations.
Numerical simulations demonstrated how the magnetization and twist overlap characteristics change between the thermal and localized phases.
We found evident differences between them, although the existence of a phase transition was not decided.
The disorder-strength dependence of the magnetization and twist overlap is robust against static noises in the interaction between spins and the local field.

Under the condition in this work, the twist overlap evaluated using the wave function at the end of the time evolution behaved similarly to that calculated using eigenstates in the middle of the energy spectrum. 
In other words, the twist overlap after time evolution can provide information on the properties of eigenstates beyond the memory effect.
The twist overlap is easily obtained from the measurement of qubits in a quantum device.
Although this work assumes an ideal quantum device, the results suggest that the twist overlap is a promising probe for detecting MBL in quantum computing approaches.



\bibliography{mbl}

\end{document}